\documentclass[12pt,onecolumn,floatfix,altaffilletter,superscriptaddress,
tightenlines,showpacs,showkeys,preprintnumbers,nofootinbib]{revtex4-1}
\pdfoutput=1

\usepackage[lmargin=80pt, rmargin=80pt,bmargin=100pt,tmargin=100pt]{geometry}
\usepackage[colorlinks=true,citecolor=blue,linkcolor=blue,breaklinks=true]{hyperref}
\usepackage{amsmath,amssymb}
\usepackage{epsfig} 
\usepackage{graphicx}        
\usepackage{url}
\usepackage{color}
\usepackage{multirow}
\usepackage{placeins}
\usepackage[dvipsnames]{xcolor}
\usepackage{braket}
\usepackage{float}
\usepackage{slashed}
 
\hypersetup{
  colorlinks=true,
  linkcolor=red,
  filecolor=magenta,   
  urlcolor=blue,
  citecolor=blue,
}

\allowdisplaybreaks

\bibpunct{[}{]}{,}{n}{}{,}

\newcommand{\rs}[1]{\textcolor{red}{}}

\makeatletter
\@addtoreset{equation}{section}

\def\thesection{\Roman{section}}

\begin{document}

\title{A radiative Type-II seesaw model with broken Symmetry Ansatz}
\author{Satyabrata Datta}
\email{satyabrata.datta@saha.ac.in}
\affiliation{Saha Institute of Nuclear Physics, HBNI, 1/AF Bidhannagar, Kolkata 700064, India}
\author{Ambar Ghosal}
\email{ambar.ghosal@saha.ac.in}
\affiliation{Saha Institute of Nuclear Physics, HBNI, 1/AF Bidhannagar,
Kolkata 700064, India}

\begin{abstract}
		Parametrization of the neutrino mass matrix in terms of well-known measured quantities is an attractive way to obtain a phenomenologically viable form. We propose a model of neutrino mass matrix based on type-II seesaw mechanism adhering the concept of badly broken symmetry. Two of the mixing angles are coming out as $\theta_{23}\sim46.08^{\circ}$, $\theta_{13}\sim8.69^{\circ}$. However, to accommodate the other oscillation parameters, we further add an extra doublet and the neutrino masses are generated through the Zee mechanism at the one loop level, and the modified model can admit all the extant data for a suitable choice of model parameters. 
\end{abstract}
	\maketitle
	\section{Introduction}
	There are different propositions to build a phenomenologically viable neutrino mass matrix following various mixing schemes \cite{Mohapatra:2006gs}. 	
	Prior to the fact that $\theta_{13}\neq0$, the Tribimaximal mixing \cite{Harrison:2002er,Xing:2002sw,Marzocca:2011dh} has been widely accepted as a correct description of the neutrino mixing. There are several models invoking different flavour symmetries to obtain the neutrino mass matrix which reconciles with this mixing pattern. After the measurement of nonzero $\theta_{13}$, this mixing has been modified in different ways considering this scheme as a leading order prediction \cite{Chen:2019fgb}. There are several other mixing schemes namely, Trimaximal mixing \cite{Lam:2006wm,Albright:2008rp,Albright:2010ap,He:2011gb,Novichkov:2018yse,Samanta:2018hqm}, Cobimaximal mixing \cite{Ma:2015fpa,Grimus:2003yn,Samanta:2017kce,Sinha:2017rjj}, Bilarge mixing \cite{Boucenna:2012xb} etc. to address nonzero $\theta_{13}$ as well as other two mixing angles within the experimental ranges. All these mixing schemes are invoked in models with different flavour symmetries to obtain the most elusive structure of the neutrino mass matrix. In this regard, we attempt to parametrize the neutrino mass matrix in terms of some experimentally known quantities. Following we try to write down the neutrino mass matrix in terms of some function of the charged lepton masses and the sum of the three electroweak neutrino masses($\sum m_{\nu_i}$). One of the possible ways to correlate the neutrino mass matrix in terms of the charged lepton masses is through the invocation of some GUT models. However, in the present work, we adopt a different approach, widely investigated earlier, is due to the assumption of broken symmetry ansatz. To demonstrate our proposed texture, we consider a type-II seesaw model. One of us has been investigated \cite{Koide:2000jm,Ghosal:2000wg} the idea of badly broken symmetry in the context of the Zee model \cite{Zee:1980ai,Smirnov:1996bv,Ghosal:2001ep} where the model leads to bimaximal mixing pattern. The crucial difference between the Zee model and the present one is that in the Zee model the diagonal elements are zero due to $SU(2)$ antisymmetry, whereas in the present case the diagonal elements are all nonzero.\\
	In the present work, we consider a model based on type-II seesaw mechanism with explicitly broken lepton number adhering ansatz of the badly broken symmetry or approximate symmetry. Inclusion of badly broken symmetry enables us to write the neutrino mass matrix in terms of the charged lepton masses and a real free parameter which can be fitted with $\sum m_{\nu_i}$. The most interesting feature of the obtained neutrino mass matrix is that the two mixing angles naturally come out as $\theta_{23}=46.08^{\circ}$, $\theta_{13}=8.69^{\circ}$, which are well within the 3$\sigma\:(99.7\% \:\text{C.L.})$ experimental ranges. However, the model does not admit all other oscillation data within the 3$\sigma\:(99.7\% \:\text{C.L.})$ experimental limits. Thus we further modified the model by adding an extra doublet Higgs, through which neutrino mass is generated at the one loop level in addition to the tree level. We further want to emphasize that the lepton number is explicitly broken in our model. Our numerical estimation shows that the tree level and the one loop level contributions are almost of the same order and this is admissible since we are not considering any perturbation.\\
	Plan of the paper is as follows: Section \ref{s1} contains our basic model. One loop modification of the model is given in Section \ref{s2}. Section \ref{s3} contains our concluding summary. Detailed calculation of the matrix elements needed through the inclusion of badly broken symmetry is given in Appendix \ref{app Appendix A}. Masses and mixing angles of the tree level neutrino mass matrix are given in Appendix \ref{app Appendix B}.

	\section{Proposed Model}\label{s1}
	We consider a model based on type-II seesaw mechanism adhering the idea of badly broken symmetry or approximate symmetry \cite{Oneda:1991wz}. The philosophy of badly broken symmetry is that some internal symmetries of a transition matrix elements become exact in the large value of a kinematical parameter, following the Goldberger-Treiman relation \cite{Goldberger:1958tr}. Earlier it has been studied in the context of $SU(3)$, chiral $SU(3)\times SU(3)$ groups \cite{Divakaran:1968zfa} in the context of hadronic and leptonic currents. In the present work, we invoke this idea in the context of a model based on type-II seesaw mechanism. Basically, in this approach, we consider the symmetry breaking effect is proportional to the masses of the charged leptons and the same symmetry breaking parameter is also responsible for the neutrino sector. \\ 
	Consider the Lagrangian of a type-II seesaw model as
	\begin{equation}\label{eqn 1}
	\begin{aligned}
	{} &\mathcal{L}=f_{ij}\overline{(l_{iL})^c}l_{jL}\Delta +y_{ij}\overline{e}_{iL}e_{jR}\phi_1+h.c.\\
	& = f_{ij}\left( \overline{(\nu_{iL})^c}\nu_{jL}\Delta^0+\overline{(\nu_{iL})^c} e_{jL} \Delta^++\overline{(e_{iL})^c} \nu_{jL}\Delta^++\overline{(e_{iL})^c}e_{jL}\Delta^{++}\right) + y_{ij}\overline{e}_{iL}e_{jR}\left\langle \phi_1^0\right\rangle +h.c.
	\end{aligned}
	\end{equation}
	where $i,j$ are the generation indices ($i,j=1,2,3$).\\
	Furthermore, we consider a strong binding between the leptons to form a condensate as $\left<\overline{e_i}e_j\right>$ and $\left<\overline{(l_{iL})^c}l_{jL}\right>$ at a very high scale above the electroweak energy scale. However, those condensates are not invariant under $SU(2)_L\times U(1)_Y$. Hence we propose an $SU(3)_H$ horizontal symmetry \cite{akama.terazawa76,Maehara:1978ts,Wilczek:1978xi,Davidson:1979wr} at a very high scale under which lepton doublets and right handed charged lepton($e_{iR}$) are forming a triplet as $l_{iL}(3)$, $e_{iR}(3)$. We further consider such symmetry is badly broken in the flavour space and the amount of symmetry breaking is proportional to the mass of the charged leptons. We admit our model is not very explicit at this stage as we do not touch the origin of the breaking of such $	SU(3)_H$ symmetry in the present work. In the present work this is our basic assumption.\\
	Now, the leptons are considered in a triplet representation of the $SU(3)_H$ group and the condensate $\left<\overline{e_i}e_j\right>$ is a component of $3\times3^*=1+8$ of the same group and the condensate  $\left<\overline{(l_{iL})^c}l_{jL}\right>$ is a component of $3\times3=6+3^*$. Again, we want to point out that the $\phi_1$ and $\Delta$ fields are not affected by our ansatz.\\
	We assume the magnitude of the matrix element in the limit $|p|\rightarrow\infty$ is given by
	\begin{equation}\label{eqn 2}
	m_{e_{ij}}\left<e_i(p)|\overline{e}_ie_j|e_j(p)\right>\propto \delta_{ij}k_i^2
	\end{equation}
	essentially which we can write as 
	\begin{equation}\label{eqn 3}
	m_{e_i}\left<e_i(p)|\overline{e}_ie_i|e_i(p)\right>=k_i^2\times Constant.
	\end{equation}
	We have removed the double subscript of $m_{e_{ii}}$ by $m_{e_i}$, 
	where $m_{e_i}=y_i{\left<\phi_1^0\right>}$. Evaluating the above matrix element given in the l.h.s. of eqn. (\ref{eqn 3}) we get
	\begin{equation}\label{eqn 4}
	\frac{1}{(2\pi)^3}\frac{m_{e_{i}}^2}{E_{e_{i}}}\overline{u}_{e_{i}}(p)u_{e_{i}}(p)=k_i^2\times Constant
	\end{equation}
	with the normalization condition $\overline{u}_{e_{i}}(p)u_{e_{i}}(p)=1$ and in the $|p|\rightarrow \infty$ limit eqn. (\ref{eqn 4}) becomes
	\begin{equation}\label{eqn 5}
	{m_{e_{i}}}^2=k_i^2\times Constant
	\end{equation}
	where $|p|$ is included in the constant.\\
	Next we consider the neutrino part of the Lagrangian and we consider the symmetry breaking as 
	\begin{equation}\label{eqn 6}
	f_{ij}\overline{(l_{iL})^c}l_{jL}\propto \sqrt{k_i k_j}.
	\end{equation}
	Explicitly we can write as
	\begin{equation}\label{eqn 7}
	\begin{aligned}
	{} & f_{ij}\left\lbrace  \left<\overline{\nu_{iL}}(p)|\overline{(\nu_{iL})^c}\nu_{jL}|\nu_{j}(p)\right>+\left<\overline{\nu_{iL}}(p)|\overline{(\nu_{iL})^c} e_{jL}|e_j(p)\right>+\right. \\
	&\left.\left<\overline{e_{iL}}(p)|\overline{(e_{iL})^c} \nu_{jL}|\nu_j(p)\right>  +\left<\overline{e_{iL}}(p)|\overline{(e_{iL})^c}e_{jL}|e_j(p)\right> \right\rbrace + i\rightarrow j = \sqrt{k_ik_j}K
	\end{aligned}
	\end{equation}
	where `$K$' is a dimensionless parameter. \\
	Evaluating l.h.s. of eqn. (\ref{eqn 7}), we get
	\begin{equation}\label{eqn 8}
	\begin{aligned}
	{} & f_{ij}\left\lbrace m_{\nu_i}+m_{e_j}+m_{e_i}+m_{e_j}+i\rightarrow j\right\rbrace \\
	& =f_{ij}.3(m_{e_j}+m_{e_i})
	\end{aligned}
	\end{equation}
	where we have neglected the neutrino masses compared to the charged lepton masses.
	Thus, we get from eqn. (\ref{eqn 8}) as
	\begin{equation}\label{eqn 9}
	3f_{ij}(m_{e_j}+m_{e_i})=\sqrt{m_{e_i}m_{e_j}}K
	\end{equation}
	which gives 
	\begin{equation}\label{eqn 10}
	f_{ij}=\dfrac{\sqrt{m_{e_i}m_{e_j}}K}{ 3(m_{e_j}+m_{e_i})}.
	\end{equation}
	Explicit calculation of the evaluation of the above matrix elements is given in the Appendix \ref{app Appendix A}. \\	
	The above expression of $f_{ij}$ leads to the following tree-level mass matrix of neutrino as
	\begin{equation}\label{eqn 11}
	m_{\nu}=\begin{pmatrix} \dfrac{K}{6} & \dfrac{K}{3}\sqrt{\dfrac{m_e}{m_\mu}}  & \dfrac{K}{3}\sqrt{\dfrac{m_e}{m_\tau}} \\ \dfrac{K}{3}\sqrt{\dfrac{m_e}{m_\mu}} & \dfrac{K}{6} & \dfrac{K}{3}\sqrt{\dfrac{m_\mu}{m_\tau}} \\\dfrac{K}{3}\sqrt{\dfrac{m_e}{m_\tau}} & \dfrac{K}{3}\sqrt{\dfrac{m_\mu}{m_\tau}} & \dfrac{K}{6}  \end{pmatrix} \left\langle \Delta^0\right\rangle. \\
	\end{equation}
	The masses and mixing angles obtained from the above matrix is given in Appendix \ref{app Appendix B}. First of all, it is to be noted that all the three mixing angles are fixed and are independent of the parameter $K^\prime$, where $K^\prime=K \left\langle \Delta^0\right\rangle$. The parameter $K^\prime$ is restricted due to the cosmological observational value of $\sum m_{\nu_i}$, and for the value of $\sum m_{\nu_i}=0.17\:$eV the bound on $K^\prime$ comes out as $K^\prime=0.34 \:$eV. The value of $ \left\langle \Delta^0\right\rangle$ is restricted most stringently from the decay of $\mu\rightarrow eee$ and is given by $ \left\langle \Delta^0\right\rangle>3.1\times 10^{-9}\:$GeV for $m_{H^{++}}$(mass of the doubly charged Higgs)$=600 $ GeV\cite{Antusch:2018svb}. The parameter $K$ is only restricted due to the perturbative unitarity bound as $K^2/4\pi<1$. Nevertheless, it is always possible to fit the product parameter $K \left\langle \Delta^0\right\rangle$ at the required range. Furthermore the value of $\left|(m_\nu) \right|_{11} $ is also below the experimental value $\left|(m_\nu) \right|_{11}\leq 0.061\:$eV as given in KamLAND-Zen and EXO-200 experiments \cite{KamLAND-Zen:2016pfg,Albert:2017owj,Vergados:2012xy,Bilenky:2012qi}. Moreover, the two mixing angles are come out as $\theta_{13}\sim 8.69^\circ$ and $\theta_{23}\sim 46.08^\circ$ which are well inside the 3$\sigma\:(99.7\% \:\text{C.L.})$ experimental limits. It is interesting to note that the above mass matrix slightly deviates from the exact mu-tau symmetric texture. The deviation is caused due to the difference between the mu and tau lepton masses following the proposed ansatz of badly broken symmetry. Let us point out the major shortcomings of the proposed texture: (1) $\theta_{12}$ value is coming out too low, (2) the mass squared differences $\Delta m_{21}^2$ and $\Delta m_{23}^2$ are also outside the 3$\sigma\:(99.7\% \:\text{C.L.})$ experimental ranges. Thus in order to accommodate all the oscillation data, we have to modify the above model.

	\section{Extended Model}\label{s2}
\begin{figure}
		\includegraphics[scale=.5]{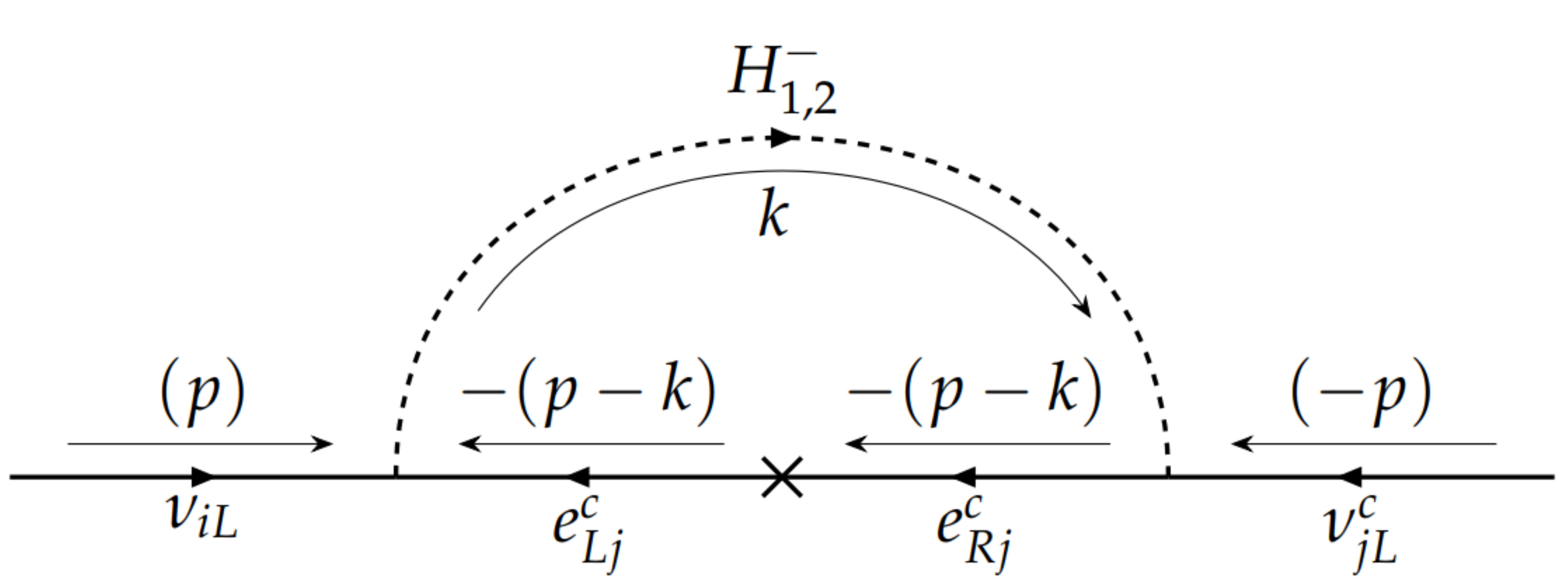}
		\caption{One-loop neutrino mass generation.}\label{loop}
	\end{figure}

	We modify the above model through the introduction of another Higgs doublet $\phi_2$ so that the masses can be generated at the one-loop due to the Zee mechanism.\\
	The relevant part of the Lagrangian is given by
	\begin{equation}\label{eqn 12}
	\mathcal{L}=2f_{ij}\overline{(e_{iL})^c}\nu_{jL}\Delta^++\frac{y_{mn}}{\sqrt{2}}\left[ \overline{(e_{Lm})^c}(e_{nR})^c\phi_2^0-\overline{(\nu_{Lm})^c}(e_{Lm})^c\phi_2^-\right]+c_{12}\phi_1^Ti\tau_2 \phi_2 \Delta 
	\end{equation}
	where $c_{12}$ is the new coupling constant.\\
	We consider the mixing between the charged scalars as
	\begin{equation}\label{eqn 13}
	\begin{aligned}
	{}& \phi_1^+=\cos\beta\: \phi^++\sin\beta \:{\phi^\prime}^+\\
	&  \phi_2^+=-\sin\beta \:\phi^++\cos\beta \:{\phi^\prime}^+
	\end{aligned}
	\end{equation}
	with $\tan\beta=v_1/v_2$, where $v_1$,$v_2$ are given by $\left\langle\phi_1^0 \right\rangle =v_1\:,\:\left\langle \phi_2^0\right\rangle =v_2$ and $v/\sqrt{2}=\sqrt{v_1^2+v_2^2}=174 \:$GeV.
	Further we consider the mixing between the $\Delta^+$ and $\phi^+$ as 
	\begin{equation}\label{eqn 14}
	\begin{aligned}
	{} & \Delta^+=\cos\alpha \:H_1^++\sin\alpha\:H_2^+\\
	& \phi^+=-\sin\alpha \:H_1^++\cos\alpha \:H_2^+
	\end{aligned}
	\end{equation}
	where  $H_1^+$ and $H_2^+$ are two remaining charged scalars defined in their mass eigenstates with the value of the mixing angle as 
	\begin{equation}\label{eqn 15}
	\begin{aligned}
	{} & \tan2\alpha=\dfrac{2\sqrt{2}c_{12}M_{W}g^{-1}}{m_\Delta^2-m_\phi^2}\\
	& =\dfrac{2\sqrt{2}c_{12}M_{W}g^{-1}}{\sqrt{(M_1^2-M_2^2)^2-(2\sqrt{2}c_{12}M_W g^{-1})^2}}
	\end{aligned}
	\end{equation}
	where $M_{1,2}$ are the masses of the charged Higgs scalars $H_1^+$ and $H_2^+$ respectively, $M_W$ is the mass of the $W^+$ boson \& $g$ is the $SU(2)_L$ coupling constant.
	Neutrino masses are generated through the charged Higgs scalars at the one-loop level as depicted in FIG.\ref{loop}. 
	
	Evaluating FIG.\ref{loop}. we get 
	\begin{equation}\label{eqn 16}
	M_{ij}=\left( \dfrac{(m_{e_i}^2+m_{e_j}^2)f_{ij}\sin{2\alpha}\sin{\beta}}{\sqrt{2}v_2}\right) \dfrac{1}{16\pi^2}\ln\left( \dfrac{M_2^2}{M_1^2}\right) 
	\end{equation}
	where $m_{e_i}$, $m_{e_j}$ are the masses of the charged leptons.\\
	The neutrino mass matrix comes out after modification of the model as
	\begin{equation}\label{eqn 17}
	m_{\nu}=\begin{pmatrix} \dfrac{K^\prime}{6}+M_{11}& \dfrac{K^\prime}{3}\sqrt{\dfrac{m_e}{m_\mu}}+M_{12}  & \dfrac{K^\prime}{3}\sqrt{\dfrac{m_e}{m_\tau}}+M_{13} \\ \dfrac{K^\prime}{3}\sqrt{\dfrac{m_e}{m_\mu}}+M_{12} & \dfrac{K^\prime}{6}+M_{22} & \dfrac{K^\prime}{3}\sqrt{\dfrac{m_\mu}{m_\tau}}+M_{23} \\\dfrac{K^\prime}{3}\sqrt{\dfrac{m_e}{m_\tau}}+M_{13} & \dfrac{K^\prime}{3}\sqrt{\dfrac{m_\mu}{m_\tau}}+M_{23} & \dfrac{K^\prime}{6}+M_{33}  \end{pmatrix}.
	\end{equation}
	
\begin{table}[H]
	\begin{center}
		\caption{Neutrino oscillation parameters used in the analysis (inclusive of SK data)\cite{Esteban:2018azc}} \label{t1}
		\vspace{2mm}\label{oscx}
		\begin{tabular}{|c|c|c|c|c|c|}
			\hline
			\hline
			${\rm Parameter}$&$\theta_{12}$&$\theta_{23}$ &$\theta_{13}$ &$ \Delta
			m_{21}^2$&$|\Delta m_{31}^2|$\\
			&$\rm degrees$&$\rm degrees$ &$\rm degrees$ &$ 10^{-5}\rm
			(eV)^2$&$10^{-3} \rm (eV)^2$\\
			\hline
			$3\sigma\hspace{1mm}{\rm
				ranges\hspace{1mm}(NO)\hspace{1mm}}$&$31.61-36.27$&$41.1-51.3$&$8.22-8.98$&
			$6.79-8.01$&$2.44-2.62$\\
			\hline
			$3\sigma\hspace{1mm}{\rm
				ranges\hspace{1mm}(IO)\hspace{1mm}}$&$31.61-36.27$&$41.1-51.3$&$8.26-9.02$&
			$6.79-8.01$&$2.42-2.60$\\
			\hline
			${\rm Best\hspace{1mm}{\rm fit\hspace{1mm}}values\hspace{1mm}(NO)}$ &
			$33.82$ & $48.6$ &  $8.60$ &$7.39$ & $2.53$\\
			\hline
			${\rm Best\hspace{1mm}{\rm
					fit\hspace{1mm}}values\hspace{1mm}(IO)}$&$33.22$&$48.8$&$8.64$&$7.39$&$2.51$\\
			\hline
		\end{tabular}
	\end{center}
\end{table}
	We consider the experimental inputs, as described in TABLE \ref{t1} \cite{Esteban:2018azc}. 
	Neglecting $M_{11}$ term in the above mass matrix, the $M_{ij}$ parameters come out within the following ranges satisfying 3$\sigma\:(99.7\% \:\text{C.L.})$ neutrino oscillation data, $\sum m_{\nu_i}$ and $|m_\nu|_{11}$ bounds as
	\begin{equation}\label{eqn 18}
	\begin{aligned}
	{} & M_{12}\rightarrow(0.0010\: \text{to}\: 0.0020)\\
	& M_{13}\rightarrow(0.0099\: \text{to}\: 0.0105) \\
	& M_{22}\rightarrow(0.0240 \: \text{to}\: 0.0269) \\
	& M_{23}\rightarrow(0.0210 \: \text{to}\: 0.0224) \\
	& M_{33}\rightarrow(0.0179 \: \text{to}\: 0.0250).
	\end{aligned}
	\end{equation}
	\begin{figure}
		\includegraphics[scale=.6]{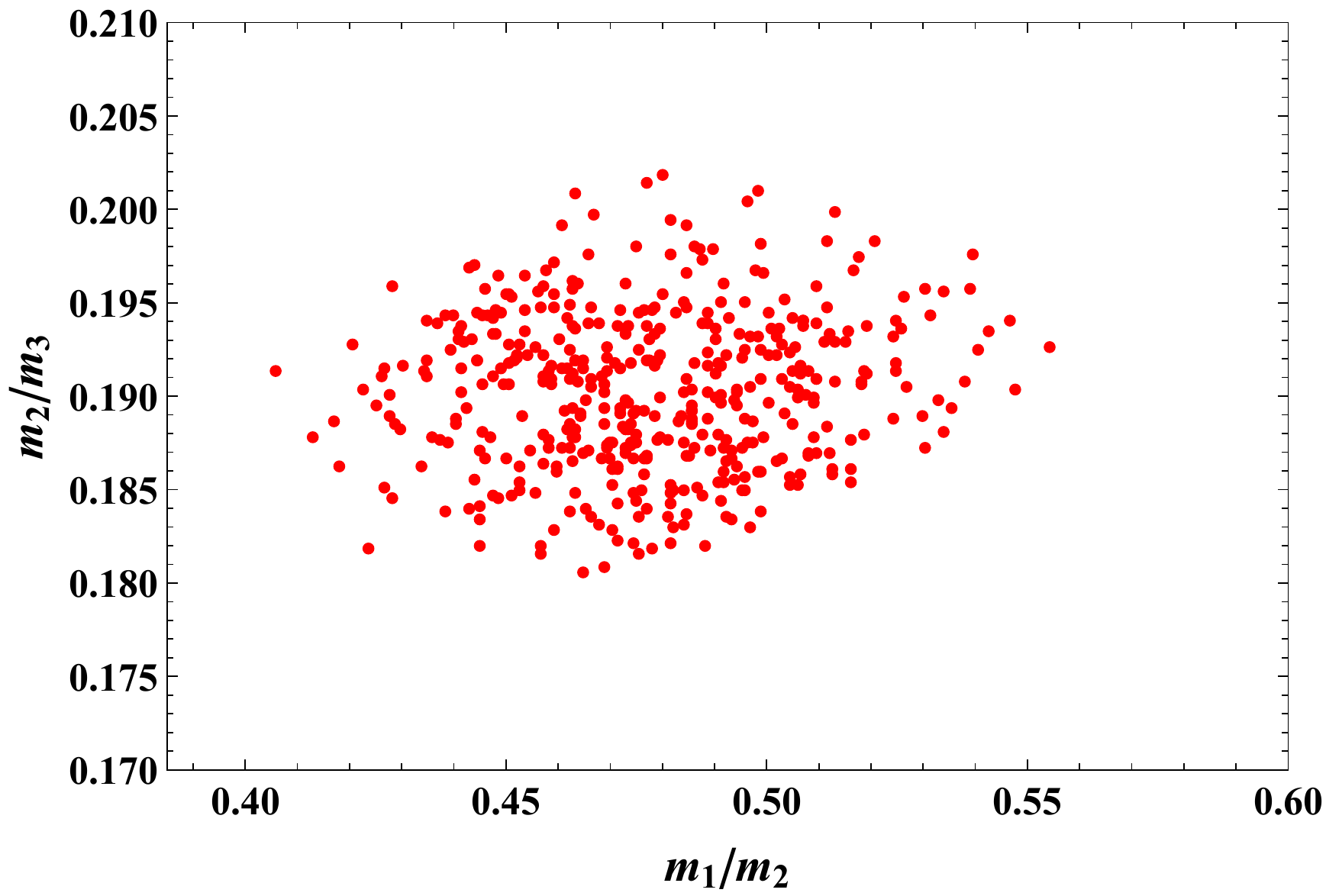}\\
		\includegraphics[scale=.6]{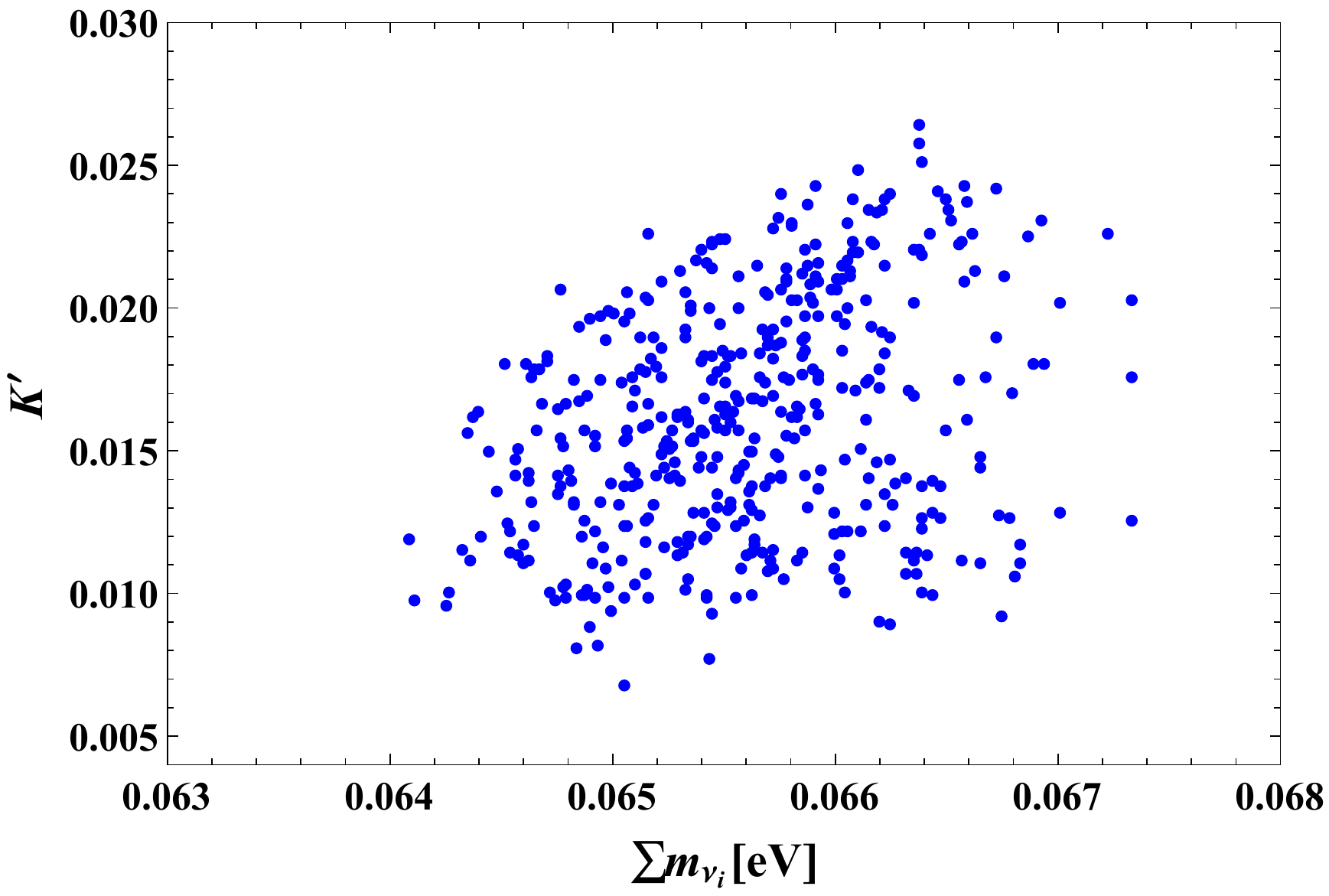}
		\caption{(a) $m_{1}/m_{2}$ vs $m_{2}/m_{3}$ and (b) variation of the parameter $K^\prime$ with $\sum m_{\nu_i}$ using $3\sigma \:(99.7\% \:\text{C.L.})$ experimental data\cite{Esteban:2018azc} given in TABLE \ref{t1}.}\label{hs}
	\end{figure}

	To find out phenomenologically viable parameter space, we have considered $2\times10^9$ random data points, and except experimental limits, no other constraints are considered. The hierarchy obtained in this case is normal and is shown in FIG.\ref{hs}(a). The allowed ranges of `$K^\prime$' parameter is also presented in FIG.\ref{hs}(b)\\
	To this end, it is worthwhile to mention some other version of such radiative neutrino mass generation. As a representative one, recently a model proposed based on\cite{Babu:2020bgz} $SU(2)_L\times SU(2)_R\times U(1)_{B-L}$ gauge group in which neutrino mass is generated at the loop level due to the inclusion of charged scalar fields. The model is not a conventional left-right symmetric model instead it includes two doublets and a charged singlet scalar fields and invokes admixture of type-I+II seesaw mechanism. The phenomenological consequence of this model is consistent with the experimental scenario, and it is worthwhile to study such type of model in view of the proposed badly broken ansatz, which will be studied elsewhere.\\
	Finally, we would also like to mention that the present model contains extra singly and doubly charged scalars and neutral Higgs fields predominantly which are decaying through leptonic channels. Collider phenomenology of this type of model already exists in the literature\cite{Antusch:2018svb,Melfo:2011nx}. LHC bounds on charged Higgs masses can be adjusted in the present model through the parameters $M_{ij}$. 
	\section{Concluding Summary}\label{s3}
	In the quest towards understanding of an elusive structure of neutrino mass matrix compatible with the extant data, in the present work, we have attempted to parametrize the neutrino mass matrix in terms of some functions of known experimental quantities in a type-II seesaw model accompanied with a badly broken $SU(3)_H$ symmetry. The model admits the value of  $\theta_{23}= 46.08^{\circ}$, and $\theta_{13}= 8.69^{\circ}$ along with $\sum m_{\nu_i}\leq0.17\:$eV and $|m_\nu|_{11}<0.061 \:$eV. Since the other oscillation data, such as $\theta_{12}$, $\Delta m_{21}^2$, $\Delta m_{32}^2$ are coming outside the present experimental limits, the present model is further modified through the introduction of another doublet scalar field. Neutrino masses are generated at the one loop level due to the Zee mechanism in addition to the tree level mass. Numerical estimation of the parameter space is also done considering $3\sigma \:(99.7\% \:\text{C.L.})$ ranges of oscillation data. Further investigation of the above model in view of CP violation and baryogenesis via leptogenesis through triplet scalars will be envisaged elsewhere. 
	\section{Acknowledgement}
	Authors acknowledge Rome Samanta for many helpful discussions.
	\medskip
	\addcontentsline{toc}{section}{References}

\begin{thebibliography}{}
		\bibitem{Mohapatra:2006gs}
		R.~N.~Mohapatra and A.~Y.~Smirnov,
		Ann. Rev. Nucl. Part. Sci. \textbf{56}, 569-628 (2006)
		doi:10.1146/annurev.nucl.56.080805.140534
		[arXiv:hep-ph/0603118 [hep-ph]].
		\bibitem{Harrison:2002er} 
		P.~F.~Harrison, D.~H.~Perkins and W.~G.~Scott,
		Phys.\ Lett.\ B {\bf 530}, 167 (2002)
		doi:10.1016/S0370-2693(02)01336-9
		[hep-ph/0202074].
		\bibitem{Xing:2002sw} 
		Z.~z.~Xing,
		Phys.\ Lett.\ B {\bf 533}, 85 (2002)
		doi:10.1016/S0370-2693(02)01649-0
		[hep-ph/0204049].
		\bibitem{Marzocca:2011dh}
		D.~Marzocca, S.~T.~Petcov, A.~Romanino and M.~Spinrath,
		JHEP \textbf{11}, 009 (2011)
		doi:10.1007/JHEP11(2011)009
		[arXiv:1108.0614 [hep-ph]].
		\bibitem{Chen:2019fgb} 
		P.~Chen, S.~Centelles Chuliá, G.~J.~Ding, R.~Srivastava and J.~W.~F.~Valle,
		Phys.\ Rev.\ D {\bf 100}, no. 5, 053001 (2019)
		doi:10.1103/PhysRevD.100.053001
		[arXiv:1905.11997 [hep-ph]].
		\bibitem{Lam:2006wm} 
		C.~S.~Lam,
		Phys.\ Rev.\ D {\bf 74}, 113004 (2006)
		doi:10.1103/PhysRevD.74.113004
		[hep-ph/0611017].
		\bibitem{Albright:2008rp} 
		C.~H.~Albright and W.~Rodejohann,
		Eur.\ Phys.\ J.\ C {\bf 62}, 599 (2009)
		doi:10.1140/epjc/s10052-009-1074-3
		[arXiv:0812.0436 [hep-ph]].
		\bibitem{Albright:2010ap} 
		C.~H.~Albright, A.~Dueck and W.~Rodejohann,
		Eur.\ Phys.\ J.\ C {\bf 70}, 1099 (2010)
		doi:10.1140/epjc/s10052-010-1492-2
		[arXiv:1004.2798 [hep-ph]].
		\bibitem{He:2011gb} 
		X.~G.~He and A.~Zee,
		Phys.\ Rev.\ D {\bf 84}, 053004 (2011)
		doi:10.1103/PhysRevD.84.053004
		[arXiv:1106.4359 [hep-ph]].
		\bibitem{Novichkov:2018yse}
		P.~P.~Novichkov, S.~T.~Petcov and M.~Tanimoto,
		Phys. Lett. B \textbf{793}, 247-258 (2019)
		doi:10.1016/j.physletb.2019.04.043
		[arXiv:1812.11289 [hep-ph]].
		\bibitem{Samanta:2018hqm} 
		R.~Samanta and M.~Chakraborty,
		JCAP {\bf 1902}, 003 (2019)
		doi:10.1088/1475-7516/2019/02/003
		[arXiv:1802.04751 [hep-ph]].
		\bibitem{Ma:2015fpa} 
		E.~Ma,
		Phys.\ Lett.\ B {\bf 752}, 198 (2016)
		doi:10.1016/j.physletb.2015.11.049
		[arXiv:1510.02501 [hep-ph]].
		\bibitem{Grimus:2003yn} 
		W.~Grimus and L.~Lavoura,
		Phys.\ Lett.\ B {\bf 579}, 113 (2004)
		doi:10.1016/j.physletb.2003.10.075
		[hep-ph/0305309].
		\bibitem{Samanta:2017kce} 
		R.~Samanta, P.~Roy and A.~Ghosal,
		JHEP {\bf 1806}, 085 (2018)
		doi:10.1007/JHEP06(2018)085
		[arXiv:1712.06555 [hep-ph]].
		\bibitem{Sinha:2017rjj} 
		R.~Sinha, R.~Samanta and A.~Ghosal,
		JHEP {\bf 1712}, 030 (2017)
		doi:10.1007/JHEP12(2017)030
		[arXiv:1706.00946 [hep-ph]].
		\bibitem{Boucenna:2012xb} 
		S.~M.~Boucenna, S.~Morisi, M.~Tortola and J.~W.~F.~Valle,
		Phys.\ Rev.\ D {\bf 86}, 051301 (2012)
		doi:10.1103/PhysRevD.86.051301
		[arXiv:1206.2555 [hep-ph]].
		\bibitem{Koide:2000jm} 
		Y.~Koide and A.~Ghosal,
		Phys.\ Rev.\ D {\bf 63}, 037301 (2001)
		doi:10.1103/PhysRevD.63.037301
		[hep-ph/0008129].
		\bibitem{Ghosal:2000wg} 
		A.~Ghosal,
		Phys.\ Rev.\ D {\bf 62}, 092001 (2000)
		doi:10.1103/PhysRevD.62.092001
		[hep-ph/0004171].
		\bibitem{Zee:1980ai} 
		A.~Zee,
		Phys.\ Lett.\  {\bf 93B}, 389 (1980)
		Erratum: [Phys.\ Lett.\  {\bf 95B}, 461 (1980)].
		doi:10.1016/0370-2693(80)90349-4, 10.1016/0370-2693(80)90193-8
		
		\bibitem{Smirnov:1996bv}
		A.~Y.~Smirnov and M.~Tanimoto,
		Phys. Rev. D \textbf{55}, 1665-1671 (1997)
		doi:10.1103/PhysRevD.55.1665
		[arXiv:hep-ph/9604370 [hep-ph]].
		\bibitem{Ghosal:2001ep}
		A.~Ghosal, Y.~Koide and H.~Fusaoka,
		Phys. Rev. D \textbf{64}, 053012 (2001)
		doi:10.1103/PhysRevD.64.053012
		[arXiv:hep-ph/0104104 [hep-ph]].
		\bibitem{Esteban:2018azc} 
		I.~Esteban, M.~C.~Gonzalez-Garcia, A.~Hernandez-Cabezudo, M.~Maltoni and T.~Schwetz,
		JHEP {\bf 1901}, 106 (2019)
		doi:10.1007/JHEP01(2019)106
		[arXiv:1811.05487 [hep-ph]].
		\bibitem{Ade:2013zuv} 
		P.~A.~R.~Ade {\it et al.} [Planck Collaboration],
		Astron.\ Astrophys.\  {\bf 571}, A16 (2014)
		doi:10.1051/0004-6361/201321591
		[arXiv:1303.5076 [astro-ph.CO]].
		\bibitem{Giusarma:2013pmn} 
		E.~Giusarma, R.~de Putter, S.~Ho and O.~Mena,
		Phys.\ Rev.\ D {\bf 88}, no. 6, 063515 (2013)
		doi:10.1103/PhysRevD.88.063515
		[arXiv:1306.5544 [astro-ph.CO]].
		\bibitem{KamLAND-Zen:2016pfg} 
		A.~Gando {\it et al.} [KamLAND-Zen Collaboration],
		Phys.\ Rev.\ Lett.\  {\bf 117}, no. 8, 082503 (2016)
		Addendum: [Phys.\ Rev.\ Lett.\  {\bf 117}, no. 10, 109903 (2016)]
		doi:10.1103/PhysRevLett.117.109903, 10.1103/PhysRevLett.117.082503
		[arXiv:1605.02889 [hep-ex]].
		\bibitem{Albert:2017owj} 
		J.~B.~Albert {\it et al.} [EXO Collaboration],
		Phys.\ Rev.\ Lett.\  {\bf 120}, no. 7, 072701 (2018)
		doi:10.1103/PhysRevLett.120.072701
		[arXiv:1707.08707 [hep-ex]].
		\bibitem{Vergados:2012xy}
		J.~D.~Vergados, H.~Ejiri and F.~Simkovic,
		Rept. Prog. Phys. \textbf{75}, 106301 (2012)
		doi:10.1088/0034-4885/75/10/106301
		[arXiv:1205.0649 [hep-ph]].
		\bibitem{Bilenky:2012qi}
		S.~M.~Bilenky and C.~Giunti,
		Mod. Phys. Lett. A \textbf{27}, 1230015 (2012)
		doi:10.1142/S0217732312300157
		[arXiv:1203.5250 [hep-ph]].
		\bibitem{Oneda:1991wz} 
		S.~Oneda and Y.~Koide,
		``Asymptotic symmetry and its implication in elementary particle physics,''
		, Singapore: World Scientific (1991) 346 p.
		\bibitem{Goldberger:1958tr} 
		M.~L.~Goldberger and S.~B.~Treiman,
		Phys.\ Rev.\  {\bf 110}, 1178 (1958).
		doi:10.1103/PhysRev.110.1178
		\bibitem{Divakaran:1968zfa} 
		P.~P.~Divakaran,
		Nucl.\ Phys.\ B {\bf 7}, 459 (1968).
		doi:10.1016/0550-3213(68)90096-5
		\bibitem{akama.terazawa76}
		K.~Akama and H.~Terazawa,
		University of Tokyo  (1976), INS-Rep 257.
		\bibitem{Maehara:1978ts} 
		T.~Maehara and T.~Yanagida,
		Prog.\ Theor.\ Phys.\  {\bf 60}, 822 (1978).
		doi:10.1143/PTP.60.822
		\bibitem{Wilczek:1978xi} 
		F.~Wilczek and A.~Zee,
		Phys.\ Rev.\ Lett.\  {\bf 42}, 421 (1979).
		doi:10.1103/PhysRevLett.42.421
		\bibitem{Davidson:1979wr} 
		A.~Davidson, M.~Koca and K.~C.~Wali,
		Phys.\ Rev.\ Lett.\  {\bf 43}, 92 (1979).
		doi:10.1103/PhysRevLett.43.92
		\bibitem{Antusch:2018svb}
		S.~Antusch, O.~Fischer, A.~Hammad and C.~Scherb,
		JHEP \textbf{02}, 157 (2019)
		doi:10.1007/JHEP02(2019)157
		[arXiv:1811.03476 [hep-ph]].
		\bibitem{Esteban:2018azc}
		I.~Esteban, M.~C.~Gonzalez-Garcia, A.~Hernandez-Cabezudo, M.~Maltoni and T.~Schwetz,
		JHEP \textbf{01}, 106 (2019)
		doi:10.1007/JHEP01(2019)106
		[arXiv:1811.05487 [hep-ph]].
		\bibitem{Babu:2020bgz}
		K.~S.~Babu and A.~Thapa,
		[arXiv:2012.13420 [hep-ph]].
		\bibitem{Melfo:2011nx}
		A.~Melfo, M.~Nemevsek, F.~Nesti, G.~Senjanovic and Y.~Zhang,
		Phys. Rev. D \textbf{85}, 055018 (2012)
		doi:10.1103/PhysRevD.85.055018
		[arXiv:1108.4416 [hep-ph]].
		
		
		
	\end{thebibliography}
	{}	
	\appendix
	\gdef\thesection{\Alph{section}}
	\makeatletter
	\renewcommand\@seccntformat[1]{Appendix \csname the#1\endcsname.\hspace{0.5em}}
	\makeatother
	
	\section{Evaluation of the matrix elements given in Section \ref{s1}.}\label{app Appendix A}
	We consider three lepton doublets ($l_1$, $l_2$, $l_3$) are as a triplet under $SU(3)_H$ so that the term $\overline{e}_ie_j$ is a component of $3\times 3^*=1+8$.\\
	We define
	\begin{equation}\label{}
	\psi(x)=\dfrac{1}{(2\pi)^{3/2}}\int d^3p \sqrt{\dfrac{m}{E}}\left[\sum_{s=1,2}^{} b_s(p)\overline{u}_s(p)e^{-ipx} +\sum_{s=1,2}^{} d_s^\dagger (p){v_s(p)}e^{ipx}\right] 
	\end{equation}
	with 
	\begin{equation}
	\left\lbrace b_s(p),b_{s^\prime}^\dagger(p^\prime)\right\rbrace =\delta_{ss^\prime}(p-p^\prime)
	\end{equation}
	and
	\begin{equation}
	\begin{aligned}{} 
	&
	\left| p,s\right\rangle =b_s^\dagger (p)\left| 0\right\rangle \\&
	\left\langle p^\prime,s^\prime|p,s\right\rangle=\delta_{ss^\prime}(p-p^\prime).
	\end{aligned}
	\end{equation}
	Thus we get
	\begin{equation}
	\begin{aligned}{}&
	\psi(x)\left| p,s\right\rangle =\dfrac{1}{(2\pi)^{3/2}}\int d^3p^\prime \sqrt{\dfrac{m}{E}}\overline{u}_s^\prime(p)e^{-ip^\prime x}b_{s^\prime}(p^\prime)b_s^\dagger(p)\left| 0\right\rangle\\
	& =\dfrac{1}{(2\pi)^{3/2}}\sqrt{\dfrac{m}{E}}\overline{u}_s(p)e^{-ipx}\left| 0\right\rangle .
	\end{aligned} 
	\end{equation}
	Therefore the matrix element comes out as
	\begin{equation}
	\begin{aligned}{}&
	\dfrac{m_{ei}}{\left<\phi^0\right>}\left\langle e_i(p)| \overline{e}_i e_j|e_j(p) \right\rangle 
	\\&
	= \dfrac{1}{(2\pi)^3}\dfrac{m_{e_i}^2}{E_i} \overline{u_i^e}(p)u_i^e(p)\dfrac{1}{\left<\phi^0\right>}
	\end{aligned}
	\end{equation}
	and using the normalization condition $\overline{u_i^e}(p)u_i^e(p)=1$ and the relation $E=|p|+m^2/2|p|$ in the Limit $|p|\rightarrow \infty$ we get
	\begin{equation}
	m_{e_i}^2=k_i^2\times Constant
	\end{equation}
	which is given in eqn. (\ref{eqn 5}).\\
	The neutrino part of the Lagrangian given in eqn. (\ref{eqn 7}) can be evaluated in the following way. The matrix element is a component of $3\times3=6+3^*$ of the $SU(3)_H$ symmetry. The first term is given by
	\begin{equation}\label{eqn 25}
	\begin{aligned}{} &
	f_{ij}\left\langle \overline{\nu_{iL}}(p)|\overline{(\nu_{iL})^c}\nu_{jL}|\nu_j(p)\right\rangle +i \leftrightarrow j \\
	& = \dfrac{f_{ij}}{(2\pi)^3}\sqrt{\dfrac{m_{\nu_j}}{E_j}}\sqrt{\dfrac{m_{\nu_i}}{E_i}}\overline{(u_{\nu_{iL}})^c}
	(u_{\nu_j})_L+i \leftrightarrow j .
	\end{aligned}
	\end{equation}
	Now in the Limit $|p|\rightarrow \infty$ we get
	\begin{equation}
	\lim\limits_{|p|\rightarrow \infty} \overline{(u_{\nu_{iL}})^c}
	(u_{\nu_j})_L =\dfrac{m_{\nu_j}}{2\sqrt{m_{\nu_i}m_{\nu_j}}}
	\end{equation}
	so we get from eqn. (\ref{eqn 25})
	\begin{equation}
	\dfrac{1}{(2\pi)^3}f_{ij} m_{\nu_i}.
	\end{equation}
	The second term
	\begin{equation}\label{eqn 28}
	\begin{aligned}{} &
	f_{ij}\left\langle \overline{\nu_{iL}}(p)|\overline{(\nu_{iL})^c}e_{jL}|e_j(p)\right\rangle +i \leftrightarrow j \\
	& = \dfrac{f_{ij}}{(2\pi)^3}\sqrt{\dfrac{m_{e_j}}{E_j}}\sqrt{\dfrac{m_{\nu_i}}{E_{\nu_i}}}\overline{(u_{\nu_{iL}})^c}
	(u_{e})_{jL}+i \leftrightarrow j .
	\end{aligned}
	\end{equation}
	In the Limit $|p|\rightarrow \infty$ we get
	\begin{equation}
	\lim\limits_{|p|\rightarrow \infty} \overline{(u_{\nu_{iL}})^c}
	(u_{e})_{jL} =\dfrac{m_{e_j}}{2\sqrt{m_{\nu_i}m_{e_j}}}
	\end{equation}
	so we get from eqn. (\ref{eqn 28}),
	\begin{equation}
	\dfrac{1}{(2\pi)^3}f_{ij} m_{e_j}.
	\end{equation}
	Similarly using the relation $\overline{(\nu_{jL})^c}e_{iL}=\overline{(e_{iL})^c}\nu_{jL}$ the third term can be evaluated as $\Rightarrow \dfrac{1}{(2\pi)^3}f_{ij} m_{e_i} $.\\
	The forth term is evaluated as 
	\begin{equation}\label{eqn 31}
	\begin{aligned}{} &
	f_{ij}\left\langle \overline{e_{iL}}(p)|\overline{(e_{iL})^c}e_{jL}|e_j(p)\right\rangle +i \leftrightarrow j \\
	& = \dfrac{f_{ij}}{(2\pi)^3}\sqrt{\dfrac{m_{e_j}}{E_j}}\sqrt{\dfrac{m_{e_i}}{E_{i}}}\overline{(u_{e_{iL}})^c}
	(u_{e})_{jL}+i \leftrightarrow j .
	\end{aligned}
	\end{equation}
	Substituting the limiting value
	\begin{equation}
	\lim\limits_{|p|\rightarrow \infty} \overline{(u_{e_{iL}})^c}
	(u_{e})_{jL} =\dfrac{m_{e_j}}{2\sqrt{m_{e_i}m_{e_j}}}.
	\end{equation}
	Thus from eqn. (\ref{eqn 31}) we get
	\begin{equation}
	\dfrac{1}{(2\pi)^3}f_{ij} m_{e_j}.
	\end{equation}
	Taking altogether, we get eqn. (\ref{eqn 8})
	\begin{equation}\label{}
	f_{ij}\left\lbrace m_{\nu_i}+m_{e_j}+m_{e_i}+m_{e_j}+i\rightarrow j\right\rbrace.
	\end{equation}
	\section{Explicit expressions of the neutrino masses and mixing angles of the proposed texture.}\label{app Appendix B}
	In this section we present explicit expressions of the neutrino masses and mixing angles of our proposed neutrino mass matrix. The expressions for mass eigenvalues(apart from constant $\left\langle\Delta^{0}\right\rangle$ factor) and mixing angles are given by
	\begin{equation}
	\begin{aligned}{}&
	m_1=\dfrac{K}{6}\left[ 1+\dfrac{4.2^{1/3}B}{m_\mu m_\tau A}+\dfrac{A}{3.2^{1/3}m_\mu m_\tau}\right], \\
	& m_2=\dfrac{K}{6}\left[ 1-\dfrac{2.2^{1/3}(1+i\sqrt{3})B}{m_\mu m_\tau A}-\dfrac{(1-i\sqrt{3})A}{6.2^{1/3}m_\mu m_\tau}\right], \\
	& m_3=\dfrac{K}{6}\left[ 1-\dfrac{2.2^{1/3}(1-i\sqrt{3})B}{m_\mu m_\tau A}-\dfrac{(1+i\sqrt{3})A}{6.2^{1/3}m_\mu m_\tau}\right], \\
	&  \theta_{23}=\tan^{-1}\left[ \dfrac{\sqrt{\left|4+4\left|\dfrac{\left( \dfrac{6 m_3}{K}-1\right) \sqrt{\dfrac{m_\tau}{m_\mu}}+2\sqrt{\dfrac{m_\mu}{m_\tau}}}{\left( \dfrac{6 m_3}{K}+1\right) }\right|^2+\left|\dfrac{\left( \dfrac{6 m_3}{K}-1\right)^2-4\dfrac{m_\mu}{m_\tau}}{\left( \dfrac{6 m_3}{K}+1\right)\sqrt{\dfrac{m_e}{m_\tau}}}\right|^2\ \right|} }{\sqrt{\left|4+4\left|\dfrac{\left( \dfrac{6 m_2}{K}-1\right) \sqrt{\dfrac{m_\tau}{m_\mu}}+2\sqrt{\dfrac{m_\mu}{m_\tau}}}{\left( \dfrac{6 m_2}{K}+1\right) }\right|^2+\left|\dfrac{\left( \dfrac{6 m_2}{K}-1\right)^2-4\dfrac{m_\mu}{m_\tau}}{\left( \dfrac{6 m_2}{K}+1\right)\sqrt{\dfrac{m_e}{m_\tau}}}\right|^2\ \right|} }\right], 
	\end{aligned}
	\end{equation}	
	
	\begin{equation}
	\begin{aligned}{}& \theta_{12}=\tan^{-1}\left[ \dfrac{2\left|\left( \dfrac{6 m_1}{K}-1\right) \sqrt{\dfrac{m_e}{m_\mu}}+2\sqrt{\dfrac{m_e}{m_\tau}}\sqrt{\dfrac{m_\mu}{m_\tau}}\right| }{\left|-4{\dfrac{m_\mu}{m_\tau}}+ \left( \dfrac{6 m_1}{K}-1\right)^2\right|} \right], \\
	& \theta_{13}=\sin^{-1}\left[\dfrac{2}{\left| \sqrt{\left|4+4\left|\dfrac{\left( \dfrac{6 m_1}{K}-1\right) \sqrt{\dfrac{m_\tau}{m_\mu}}+2\sqrt{\dfrac{m_\mu}{m_\tau}}}{\left( \dfrac{6 m_1}{K}+1\right) }\right|^2+\left|\dfrac{\left( \dfrac{6 m_1}{K}-1\right)^2-4\dfrac{m_\mu}{m_\tau}}{\left( \dfrac{6 m_1}{K}+1\right)\sqrt{\dfrac{m_e}{m_\tau}}}\right|^2\ \right|} \right| } \right] 
	\end{aligned}
	\end{equation}
	where 
	\begin{equation}
	\begin{aligned}{}&
	B=\left(m_em_\mu^2m_\tau+m_\mu^3m_\tau+m_em_\mu m_\tau^2 \right)  ,\\
	&A= \left( 432\sqrt{\dfrac{m_e}{m_\mu}}m_\mu^3\sqrt{\dfrac{m_e}{m_\tau}}\sqrt{\dfrac{m_\mu}{m_\tau}}m_\tau^3+\right. \\
	& \left. \sqrt{186624m_e^2m_\mu^6 m_\tau^4-6912(m_e m_\mu^2m_\tau+m_\mu^3 m_\tau+m_e m_\mu m_\tau^2)^3}\right)^{1/3}. \\
	\end{aligned}
	\end{equation}   
\end{document}